\documentclass[a4paper]{article}
\pdfoutput=1 % if your are submitting a pdflatex (i.e. if you have
\usepackage[margin=1in]{geometry}
\usepackage{graphicx}
\usepackage{amssymb}
\usepackage{xcolor}
\usepackage{lineno}
\usepackage{authblk}
\usepackage{hyperref}
\usepackage[british,UKenglish]{babel}
\begin{document}

\title{Measuring the atmospheric neutrino oscillation parameters and constraining the 3+1 neutrino model with ten years of ANTARES data}

%\documentclass{elsarticle}
%\journal{}
%\begin{document}
%\begin{frontmatter}
%\title{ANTARES collaboration}
\author[a]{A.~Albert}
\author[b]{M.~Andr\'e}
\author[c]{M.~Anghinolfi}
\author[d]{G.~Anton}
\author[e]{M.~Ardid}
\author[f]{J.-J.~Aubert}
\author[g]{J.~Aublin}
\author[g]{T.~Avgitas}
\author[g]{B.~Baret}
\author[h]{J.~Barrios-Mart\'{\i}}
% ...

\affil[a]{\scriptsize{Universit\'e de Strasbourg, CNRS,  IPHC UMR 7178, F-67000 Strasbourg, France}}
\affil[b]{\scriptsize{Technical University of Catalonia, Laboratory of Applied Bioacoustics, Rambla Exposici\'o, 08800 Vilanova i la Geltr\'u, Barcelona, Spain}}
\affil[c]{\scriptsize{INFN - Sezione di Genova, Via Dodecaneso 33, 16146 Genova, Italy}}
\affil[d]{\scriptsize{Friedrich-Alexander-Universit\"at Erlangen-N\"urnberg, Erlangen Centre for Astroparticle Physics, Erwin-Rommel-Str. 1, 91058 Erlangen, Germany}}
\affil[e]{\scriptsize{Institut d'Investigaci\'o per a la Gesti\'o Integrada de les Zones Costaneres (IGIC) - Universitat Polit\`ecnica de Val\`encia. C/  Paranimf 1, 46730 Gandia, Spain}}
\affil[f]{\scriptsize{Aix Marseille Univ, CNRS/IN2P3, CPPM, Marseille, France}}
\affil[g]{\scriptsize{APC, Univ Paris Diderot, CNRS/IN2P3, CEA/Irfu, Obs de Paris, Sorbonne Paris Cit\'e, France}}
\affil[h]{\scriptsize{IFIC - Instituto de F\'isica Corpuscular (CSIC - Universitat de Val\`encia) c/ Catedr\'atico Jos\'e Beltr\'an, 2 E-46980 Paterna, Valencia, Spain}}
% ...
%\end{frontmatter}
%\end{document}

\date{}
\maketitle
\newpage

\begin{abstract}
The ANTARES neutrino telescope has an energy threshold of a few tens of GeV. This allows to study the phenomenon of atmospheric muon neutrino disappearance due to neutrino oscillations. In a similar way, constraints on the 3+1 neutrino model, which foresees the existence of one sterile neutrino, can be inferred. Using data collected by the ANTARES neutrino telescope from 2007 to 2016, a new measurement of $\Delta m^2_{32}$ and $\theta_{23}$ has been performed - which is consistent with world best-fit values - and constraints on the 3+1 neutrino model have been derived. 
\end{abstract}

\section{Introduction}
\label{sec:1}
Neutrino oscillations arise from the mixing between flavour ($\nu_e, \nu_{\mu}, \nu_{\tau}$) and mass ($\nu_1, \nu_2, \nu_3$) eigenstates. The mixing parameters of the Pontecorvo-Maki-Nakagawa-Sakata matrix~\cite{MNS,P1,P2} (PMNS) and the differences between the mass eigenvalues regulate the oscillation probability.

Neutrino oscillations have been detected by a variety of experiments, studying solar %~\cite{SolNu, SNO-CC, SNO-NC} 
as well as atmospheric neutrinos, %~\cite{ SKFirstOsc, AntOsc, ICOsc, macro}, 
but also neutrinos produced from nuclear reactors %~\cite{DayaBay, RENOFirst, Double-Chooz} 
and particle accelerators. %~\cite{T2KComb, NOvA}
\textcolor{black}{For a comprehensive review see~\cite{PDG2018}}.

Atmospheric neutrinos are produced through the interaction of cosmic rays with nuclei in the Earth's atmosphere. Their flux spans many orders of magnitude in energy, from \,GeV to hundreds of \,TeV. Being isotropic to first order, it allows to investigate a large range of baselines on the Earth's surface, from $\sim$10\,km of vertically down-going to $\sim$10$^4$\,km of vertically up-going neutrinos.

\textcolor{black}{In this paper} the muon disappearance channel ($P_{\nu_{\mu}\rightarrow\nu_{\mu}}$) is studied. 
%namely the survival probabilities of muon neutrinos  and muon anti-neutrinos ($P_{\bar{\nu}_{\mu}\rightarrow\bar{\nu}_{\mu}}$). In the standard 3-flavour scenario, 
The vacuum survival probability for a muon neutrino of energy $E$ interacting at a distance $L$ from its creation point is given by:
\begin{equation}
P_{\nu_{\mu}\rightarrow\nu_{\mu}} =1-4\sum_{j>i} |U_{\mu j}|^2|U_{\mu i}|^2 \sin^2(\frac{\Delta m^2_{ji} L}{4 E})\sim 1-4|U_{\mu 3}|^2(1-|U_{\mu 3}|^2)\sin^2(\frac{\Delta m^2_{32} L}{4 E}),
\label{eq:1}
\end{equation}
\textcolor{black}{where $U_{\mu i}, U_{\mu j}$ are elements of the PMNS matrix $U$, and $\Delta m^2_{ji}=m^2_j-m^2_i$ are the mass splittings between two mass eigenstates. The rightmost term describes the ``single $\Delta m^2$ dominance'' approximation, relevant in the energy domain considered for this analysis. Here the $\nu_\mu$ survival probability depends  only on $U_{\mu 3}=\sin\theta_{23}\cos\theta_{13}$ and $\Delta m^2_{32}$.}
% is one of the elements of the PMNS matrix $U$, and $\Delta m^2_{32}=m^2_3-m^2_2$ is the mass splitting between the two mass eigenstates $\nu_2$ and $\nu_3$. The non-approximated expression would take into account also a dependence of the oscillation probability on the CP-violating phase, $\delta_{CP}$, and on $\theta_{12}$ and $\Delta m^2_{21}$. Matter effects~\cite{Wolfen,MS1,MS2} are neglected as they have no measurable impact in the energy range accessible to ANTARES.
For a vertically up-going atmospheric $\nu_{\mu}$, the first minimum of the survival probability described in Equation~\ref{eq:1} is reached at energies of $\sim$ 25\,GeV.
\textcolor{black}{The formalism given in Eq.~\ref{eq:1} is further modified by matter effects~\cite{Wolfen,MS1,MS2} as the neutrinos propagate through the Earth. Throughout the paper, oscillation probabilities are calculated with the OscProb package~\cite{OscProb} which treats matter effects for an arbitrary number of neutrino families numerically without approximations}.

The ANTARES neutrino telescope~\cite{Ant} has been designed and optimised for the exploration of the high-energy Universe by using neutrinos as cosmic probes. However, its energy threshold of about 20\,GeV is sufficient, even if at the edge, to be sensitive to the first atmospheric oscillation minimum, making also the study of neutrino oscillations possible. As neutrinos and antineutrinos are indistinguishable on an event-by-event basis in neutrino telescopes, in the following muon (electron) neutrinos are refered to the sum of contributions from both neutrinos and antineutrinos.

A previous analysis of ANTARES data, covering the data acquisition period from 2007 to 2010, represented the first study of this kind performed by a neutrino telescope, and measured the atmospheric neutrino oscillation parameters, $\Delta m^2_{32}$ and $\theta_{23}$~\cite{AntOsc}. In the present work, data collected during 10 years have been studied with a new analysis chain that also includes a more comprehensive treatment of various systematic effects.

Despite the fact that neutrino oscillation is a well established phenomenon, some observed experimental anomalies, such as the ones reported by the LSND~\cite{LSND} and MiniBooNE~\cite{MiniBooNE} collaborations, seem to indicate a deviation from the standard 3-flavour picture. These discrepancies could be partially explained by introducing in the model an additional neutrino state. However, since the number of weakly interacting families of light neutrinos is limited to three by the LEP results~\cite{LEP}, the additional neutrinos have to be sterile, i.e., they do not undergo weak interactions.

The 3+1 neutrino model foresees the existence of one sterile neutrino in addition to the three standard ones. \textcolor{black}{A choice has to be made, how to extend the mixing matrix $U$ from three to four families. In this analysis the convention from~\cite{MinosSt} (see ``supplementary materials'') is adopted: 
$U_{3+1}=R_{34}R_{24}R_{14}R_{23}R_{13}R_{12}$ where $R_{ij}$ is the rotation matrix for angle $\theta_{ij}$. If $j-i>1$, $R_{ij}$ also contains a CP-violating phase, $\delta_{ij}$.} Six new real mixing parameters have to be accounted for: three new mixing angles, $\theta_{14}$, $\theta_{24}$ and $\theta_{34}$, a new mass splitting, $\Delta m^2_{41}$, and two new phases, $\delta_{14}$ and $\delta_{24}$. \textcolor{black}{In line with other analyses of sterile neutrinos in the muon disappearance channel~\cite{MinosSt,SKSt,ICStLow}, $\theta_{14}=0$ is assumed, which also eliminates any dependency on $\delta_{14}$.}

\begin{figure}[htbp]
\centering
%\captionsetup{width=0.8\linewidth}
\includegraphics[width=0.8\linewidth]{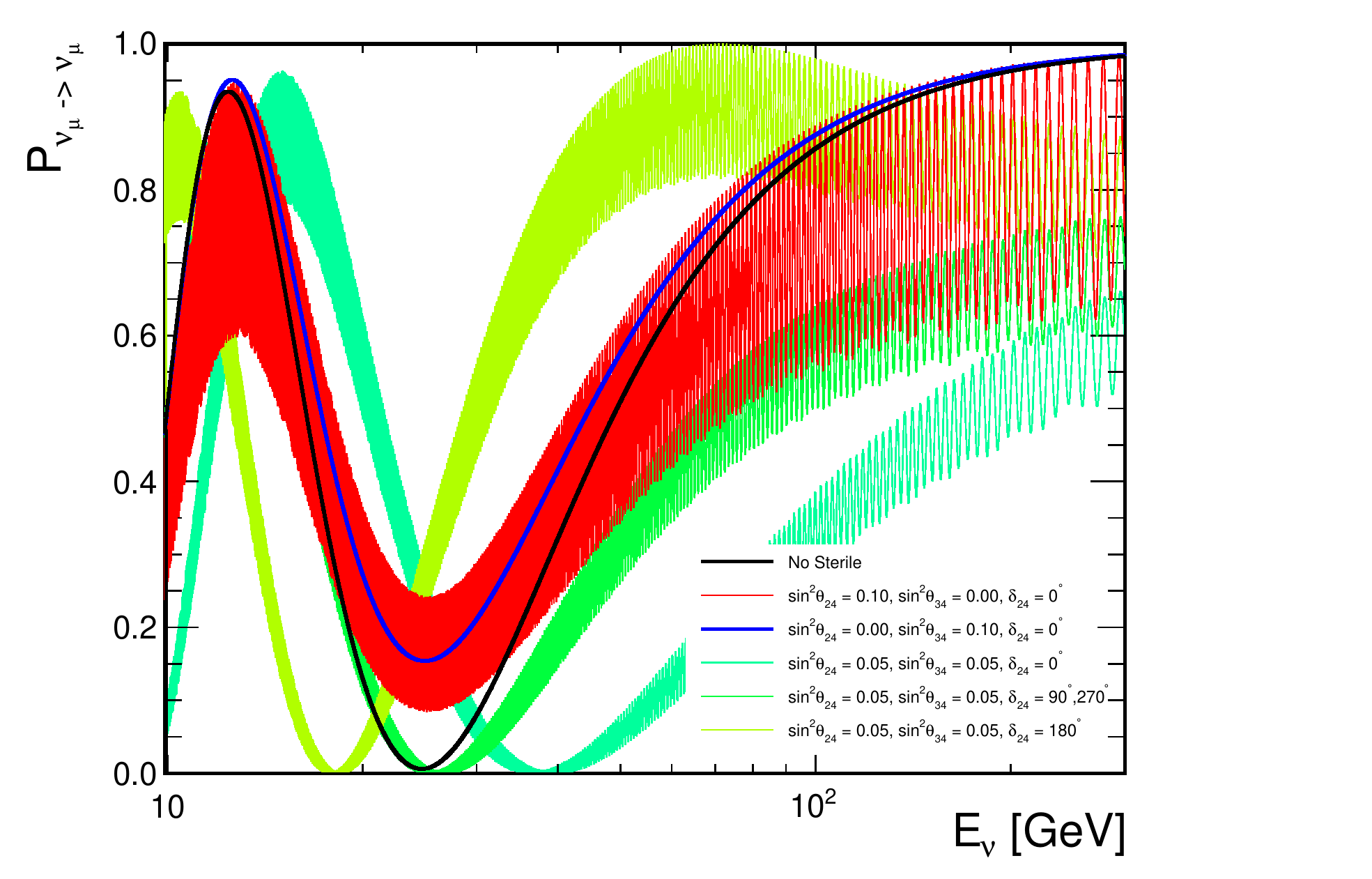}
\caption{Survival probability of vertically up-going $\nu_\mu$ as a function of neutrino energy (calculated with~\cite{OscProb}) for different values of mixing angles $\theta_{24}, \theta_{34}$ and $\delta_{24}$ with $\Delta m^2_{41}=0.5$~eV$^2$, \textcolor{black}{$\Delta m^2_{31}=2.5\cdot 10^{-3}$~eV$^2$ and $\sin^22\theta_{23}=1$.}}
\label{fig:Sterile}
\end{figure}

Even though a sterile neutrino does not interact as the active flavours, its presence would still modify the oscillation pattern of the standard neutrinos, due to the fact that the standard neutrino flavours could oscillate into these additional sterile species. In particular, for up-going $\nu_{\mu}$ in the energy range of 20-100\,GeV, non-zero values of \textcolor{black}{$U_{\mu 4}$ and $U_{\tau 4}$ with
\begin{eqnarray}
U_{\mu 4}  &=& e^{-i\delta_{24}}\sin\theta_{24},\\ 
U_{\tau 4} &=& \sin\theta_{34}\cos\theta_{24}. 
\label{Umu4}
\end{eqnarray}}
can lead to distortions in their survival probability. This is illustrated in Figure~\ref{fig:Sterile} which shows the $\nu_{\mu}$ survival probability for maximal mixing of $\theta_{23}$ and different combinations of the mixing parameters $\theta_{24},\theta_{34}$ and $\delta_{24}$. \textcolor{black}{If only $\theta_{34}$ is non-zero, the survival probability of $\nu_\mu$ with respect to the non-sterile hypothesis is only modified close to the first oscillation minimum.} The case of both $\theta_{24}$ and $\theta_{34}$ being non-zero leads instead to a significant shift of the first oscillation minimum in energy (depending on $\delta_{24}$) and modifies the event rate up to energies of few hundred GeV, easily accessible with ANTARES. The fast wiggles due to $\Delta m^2_{41}=0.5$~eV$^2$ will be smeared out by detector resolution effects, therefore no sensitivity to this parameter is expected. \textcolor{black}{The surprisingly strong effect of $\delta_{24}$ on the $\nu_\mu$ survival probability, neglected in all similar analyses so far, is further detailed in the Appendix.}

Since the effect of an additional sterile neutrino would be visible in the same energy and zenith range as the $\nu_{\mu}$ disappearance, the same analysis chain and data sample can be exploited to constrain the 3+1 neutrino model parameters. In this paper, the results of an investigation aiming to constrain the mixing angles $\theta_{24}$ and $\theta_{34}$ of the 3+1 neutrino model are also reported.

The paper is organised as follows: in Section~\ref{sec:2} the ANTARES neutrino telescope is briefly described and its detection principle is illustrated; the ANTARES data sample as well as the Monte Carlo (MC) chain are presented in Section~\ref{sec:3}, while the event reconstruction is discussed in Section~\ref{sec:4}. Section~\ref{sec:5} is dedicated to the event selection and the minimisation procedure. The results are presented in Section~\ref{sec:6}, while conclusions are given in Section~\ref{sec:7}.

\section{The ANTARES neutrino telescope}
\label{sec:2}
The ANTARES neutrino telescope is located in the Mediterranean Sea, 40\,km off the coast of Toulon, France, at a mooring depth of about 2475\,m. The detector was completed in 2008. ANTARES is composed of 12 detection lines, each one equipped with 25 storeys of 3 optical modules (OMs), except line 12 with only 20 storeys of OMs, for a total of 885 OMs. The horizontal spacing among the lines is $\sim$60\,m, while the vertical spacing between the storeys is 14.5\,m. Each OM hosts a 10-inch photomultiplier tube (PMT) from Hamamatsu~\cite{Hamamatsu}, whose axis points 45$^{\circ}$ downwards. All signals from the PMTs that pass a threshold of 0.3 single photoelectrons (hits) are digitised and sent to the shore station~\cite{Elec, DA}. The on-shore trigger system~\cite{AntTrig} performs a hit selection based on causality relations and builds events under the hypothesis that the selected hits originate from Cherenkov radiation induced by relativistic charged particles as they are produced in neutrino interactions close to the ANTARES instrumented volume.

The main sources of optical background registered by the ANTARES PMTs are represented by Cherenkov light from decay products of the radioactive isotope $^{40}\mathrm{K}$, naturally present in sea-water, by light emitted through bioluminescence by living organisms, and by energetic atmospheric muons, which can penetrate deeply under the sea and reach the detector from above.

\section{ANTARES data and Monte Carlo samples}
\label{sec:3}   
ANTARES data collected from 2007 to 2016 have been considered in the analysis. After excluding data acquired under adverse conditions, a total of 2830 days of live time has been evaluated.

The aim of the MC production is to reproduce in the most realistic way the events expected at the detector, as well as the response of the apparatus when recording these events. In order to account for changes of the environmental conditions, as well as for the different operational status of the detector and its components over time, a run-by-run MC approach is applied~\cite{RbrProc}. A typical run lasts few hours.
\textcolor{black}{Several time dependent conditions are taken from real data and applied to the run-by-run MC. First, temporarily or permanently non-operational OMs are masked in the simulation. Secondly, background light conditions, which might vary due to bioluminescence, are measured every 104~ms for each individual OM. These samples are directly used as input for the background light simulation. Thirdly, individual OM efficiencies are considered, as calculated on an approximately weekly basis from $^{40}$K coincidence rates~\cite{K40Pap}. Finally, the acoustics based position calibration, performed every few minutes, is applied. All these detailed inputs assure an authentic description of the detector response for each individual run. Remaining uncertainties are small and can be handled as global parameters which are discussed below. They are included in the analysis as systematic uncertainties.}

%Namely, the particular conditions at the time of a data run acquisition are used as input for the MC simulation of the corresponding run.

Neutrino interactions of all flavours have been simulated with the GENHEN~\cite{Genhen} package, developed inside the ANTARES Collaboration. It allows to reproduce neutrino interactions in the GeV to multi-PeV energy range. MC neutrino events can be weighted to reproduce different physical expectations. For atmospheric neutrinos with $E_{\nu}\in[20-100]$\,GeV, a MC sample almost three hundreds times larger than the data sample is available. The model by Honda et al.~\cite{Honda} for the Fr\'ejus site is used in this work.

Even though the sub-marine location of ANTARES provides a good shielding against atmospheric muons, still a large amount of them will reach the detector. The event generator used in ANTARES to simulate atmospheric muons is MUPAGE~\cite{Mupage}; the energy and angular distributions, as well as the multiplicity of muons propagating in sea water are parameterised. The contribution from this background is also evaluated from the data itself.

Particle propagation and Cherenkov light production are simulated using a GEANT-based~\cite{geant3} package~\cite{Genhen}, which takes into account all relevant physics processes and computes the probability that photons emitted by a particle reach the OM surface, producing a hit. Finally the detector response is simulated, including the digitisation and filtering of hits. At this stage a realistic optical noise is added on each OM for each data acquisition run of the detector, and the time evolution of the detector configuration is accounted for as described above. %Also the individual OM efficiencies~\cite{K40Pap} are taken into consideration.
 
\section{Event reconstruction}
\label{sec:4}
Charged-current (CC) interactions of muon neutrinos produce a muon propagating through the detector and inducing Cherenkov light. They are identified as track-like events. The event reconstruction and selection used in the analysis have been optimised to select such events. On the other hand, $\nu_{e}$ CC interactions, as well as neutral-current interactions (NC) of all flavours produce hadronic showers. In the case of $\nu_{e}$ CC interactions an electromagnetic shower is produced as well. Moreover, $\nu_\tau$ CC events can be produced as the result of $\nu_\mu\rightarrow\nu_\tau$ oscillations with and without muons in the final state. All these events constitute an additional source of background for this study.

Events have been reconstructed using two different algorithms, described in detail in~\cite{BBFit, GridFit}. In the following discussion these algorithms will be referred to as method $\mathcal{A}$ and method $\mathcal{B}$, respectively. Both are optimised for events induced by GeV-scale $\nu_{\mu}$ CC interactions. In method $\mathcal{A}$ a hit selection, based on time and spatial coincidences of hits, is applied and a $\chi^2$-fit is performed in order to find the best track. Events can have a single-line topology (SL), if all the selected hits have been recorded in the same detector line, or a multi-line topology (ML), when hits belong to OMs of different lines. Method $\mathcal{B}$ consists of a chain of fits, aimed to improve at each step the track estimation. Starting from a hit selection, a first prefit, based on a directional scan with a large number of isotropically distributed directions, is performed. The best 9 directions are used as starting points for the final likelihood ($\log \mathcal{L}$) fit.

Once the muon track has been reconstructed, its length, $L_{\mu}$, is computed. This is done, for ML events, by projecting back to the track the first and last selected hit. For SL events, since a vertex estimation is not possible due to the lack of azimuth information, the track length is estimated from the z-coordinates of the uppermost and lowermost storey which have recorded the selected hits and taking into account the reconstructed zenith angle.

The muon energy estimation is based on the fact that muons in the few-GeV energy range can be treated as minimum ionising particles, and their energy can be estimated from their track length $L_{\mu}$:
\begin{equation}
E_\mathrm{reco} = L_{\mu}\times 0.24\,\textrm{GeV/m},
\label{eq:2}
\end{equation}
where the factor 0.24\,GeV$/$m represents the energy loss of muons in sea water in the energy range of 10--100~GeV~\cite{PDG}. This quantity is used in the following as estimator for the neutrino energy.
\textcolor{black}{The energy resolution of fully contained muons is dominated by the spacing of the detector elements and is found to be around 5~GeV. For muons leaving the detector only a lower limit for their energy can be derived, corresponding to their visible length inside the instrumented volume. More details on the muon energy resolution can be found in~\cite{AntOsc}.}

\section{Analysis}
\label{sec:5}

To achieve the best sensitivity to the measurement of the oscillation parameters, a set of quality criteria has been applied. The selection of $\nu_\mu$ CC events has been optimised by performing a preliminary Monte Carlo (MC) sensitivity study, before applying the whole analysis chain to data.

The main parameter on which the selection is based is the reduced $\chi^2$ for method $\mathcal{A}$ and the $\log\mathcal{L}$ for method $\mathcal{B}$. Events reconstructed by method $\mathcal{A}$ and passing the corresponding event selection are kept. The events discarded by this procedure are further reconstructed by method $\mathcal{B}$; they are kept in the analysed sample if the corresponding selection criteria are passed. Only events which are reconstructed as up-going are used in the following. A minimum number of five storeys with selected hits is required, in order to minimise the background induced by atmospheric muons.

\begin{figure}[htbp]
\centering
\includegraphics[width=0.8\linewidth]{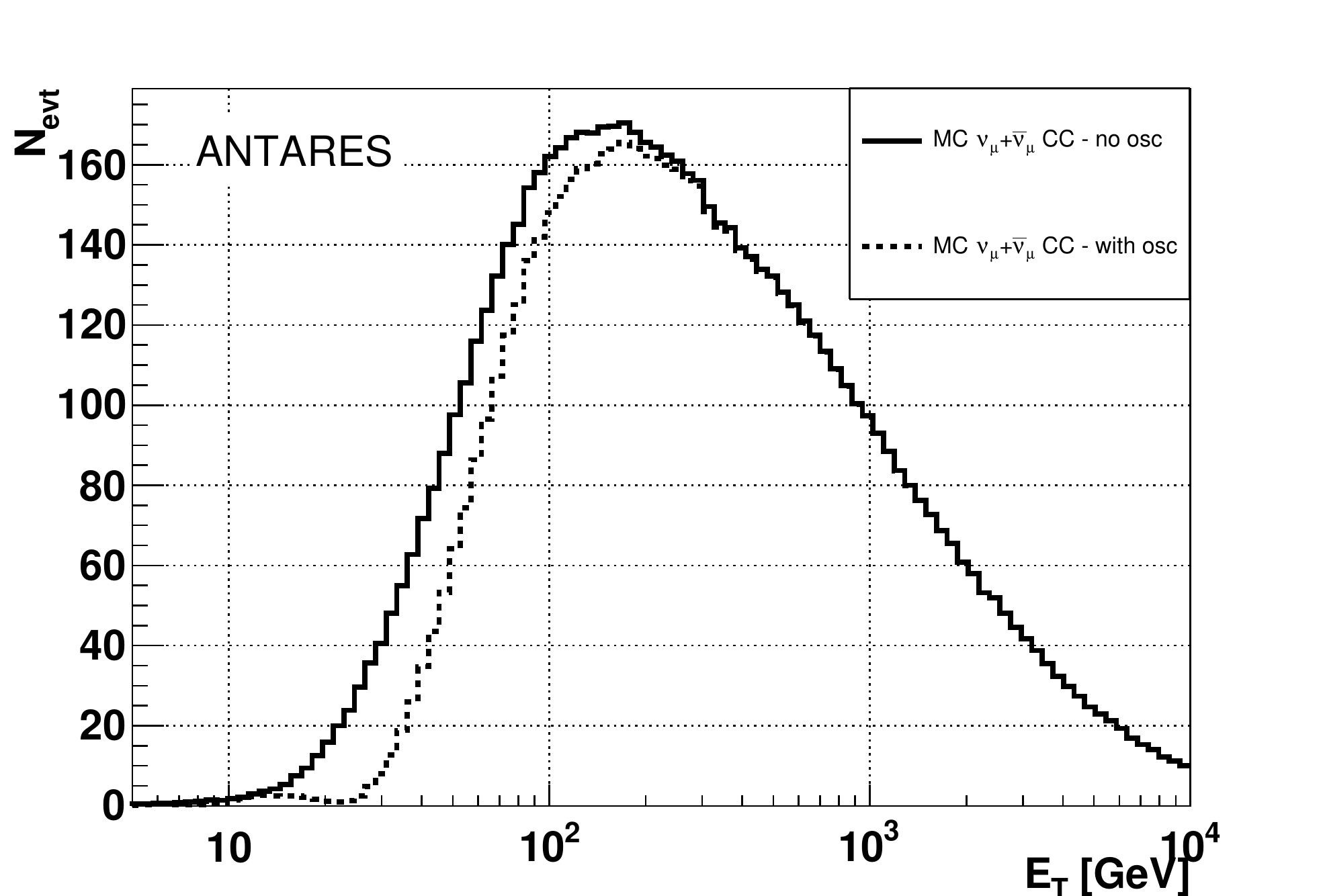}
\caption{MC neutrino energy, $\mathrm{E_T}$, for selected $\nu_{\mu}$ and $\bar{\nu}_{\mu}$ CC events: assuming no oscillations (solid line) and a 2-flavour oscillation scenario with maximal mixing and $\Delta m^2_{32}=2.46\times10^{-3}$\,eV$^2$ (dashed line).}
\label{fig:1}
\end{figure}

In Figure~\ref{fig:1} the distribution of the MC true neutrino energy, $\mathrm{E_T}$, for selected $\nu_{\mu}$ CC events is shown.
For the histogram with the solid line no neutrino oscillations are assumed, while the dashed one refers to a 2-flavour oscillation scenario with maximal mixing and $\Delta m^2_{32}=2.46\times10^{-3}$\,eV$^2$.
As can be seen, atmospheric neutrino oscillations affect the expected event distribution for $\mathrm{E_T} \lesssim 100$~GeV. About $7590$ well-reconstructed $\nu_\mu$ CC events are expected in a live time of 2830 days when oscillations are neglected. Roughly one half of these events are reconstructed with method $\mathcal{A}$ (ML), while methods $\mathcal{A}$ (SL) and $\mathcal{B}$ both contribute with approximately 25\% to this event sample. Further, $\sim$40 $\nu_e$ CC events are selected. Oscillations reduce the number of expected events by $\sim$720 events. This reduction is dominantly seen in the $\mathcal{A}$ (SL) sample ($\sim 60\%$) which contains the lowest energetic and most vertical events, while the other two reconstruction methods contribute each about $20\%$. $\nu_\tau$ CC events reduce this oscillation signal by $\sim 20$~events, taking into account the energy-dependent cross section ratio $\sigma(\nu_\tau~\mathrm{CC})/\sigma(\nu_\mu~\mathrm{CC})$ (about 0.5 at 25 GeV), the 17\% branching ratio of the muonic $\tau$ decay and the resulting soft spectrum of the produced muons.

\begin{figure}[htbp]
\centering
\includegraphics[width=0.8\linewidth]{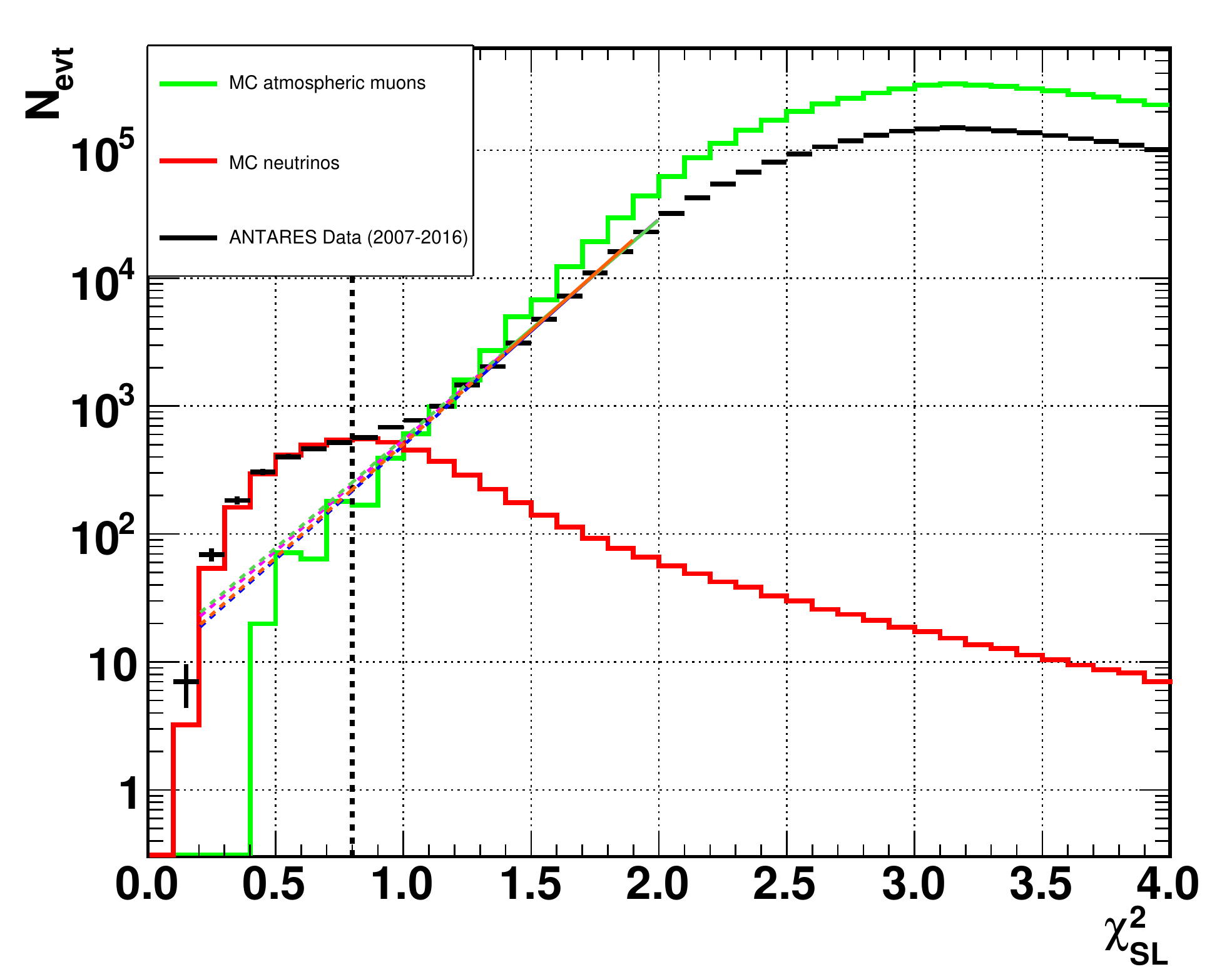}
\caption{Distribution of reduced $\chi^2_{\mathrm{SL}}$ values for events which have been reconstructed by method $\mathcal{A}$ (SL). Data (black crosses) with error bars indicating the statistical uncertainty are shown together with MC neutrino events (red line) and MC atmospheric muons (green line). The dashed black line at $\chi^2_{\mathrm{SL}}=0.8$ indicates the value of the applied cut on this parameter. The fitted functions used to estimate the background of atmospheric muons are shown as well (solid coloured lines), together with their extrapolation into the signal region left to the cut value (dashed coloured lines, see text for details).}
\label{fig:2}
\end{figure}

Figure~\ref{fig:2} shows the distribution of the reduced $\chi_{\mathrm{SL}}^2$ for method $\mathcal{A}$ (SL) events where data are compared to simulated atmospheric neutrinos and background atmospheric muons. While the MC reproduces quite well the data in the signal region dominated by the neutrino signal, a disagreement between the MC expectation and data is visible for larger $\chi^2_{\mathrm{SL}}$. 
\textcolor{black}{Both data and MC follow an exponential law in this region, but with different slopes.}
For this reason, the number of background atmospheric muons in the signal region has been determined from data itself. The distribution in Figure~\ref{fig:2} has been parameterised in the region dominated by atmospheric muons ($\chi^2_{\mathrm{SL}}>0.8$) with four different exponential fits by varying the fit range. Each fit has been extrapolated into the signal region, and its corresponding integral has been computed. The mean of these integrals has been used to estimate the number of atmospheric muon background, and its uncertainty has been computed from the errors on the fitted function parameters. Summing up the results of this method for events that have been reconstructed by method $\mathcal{A}$ (SL and ML) and method $\mathcal{B}$, and combining the corresponding errors in quadrature, a total background of $740\pm 120$ atmospheric muons has been determined. This value is subsequently used as a Gaussian prior mean value and uncertainty in the minimisation procedure. The energy and direction distribution of the atmospheric muon background has been, instead, estimated directly from MC.

\begin{figure}[htbp]
\begin{minipage}[h!]{0.49\linewidth}
\centering
\includegraphics[width=\linewidth]{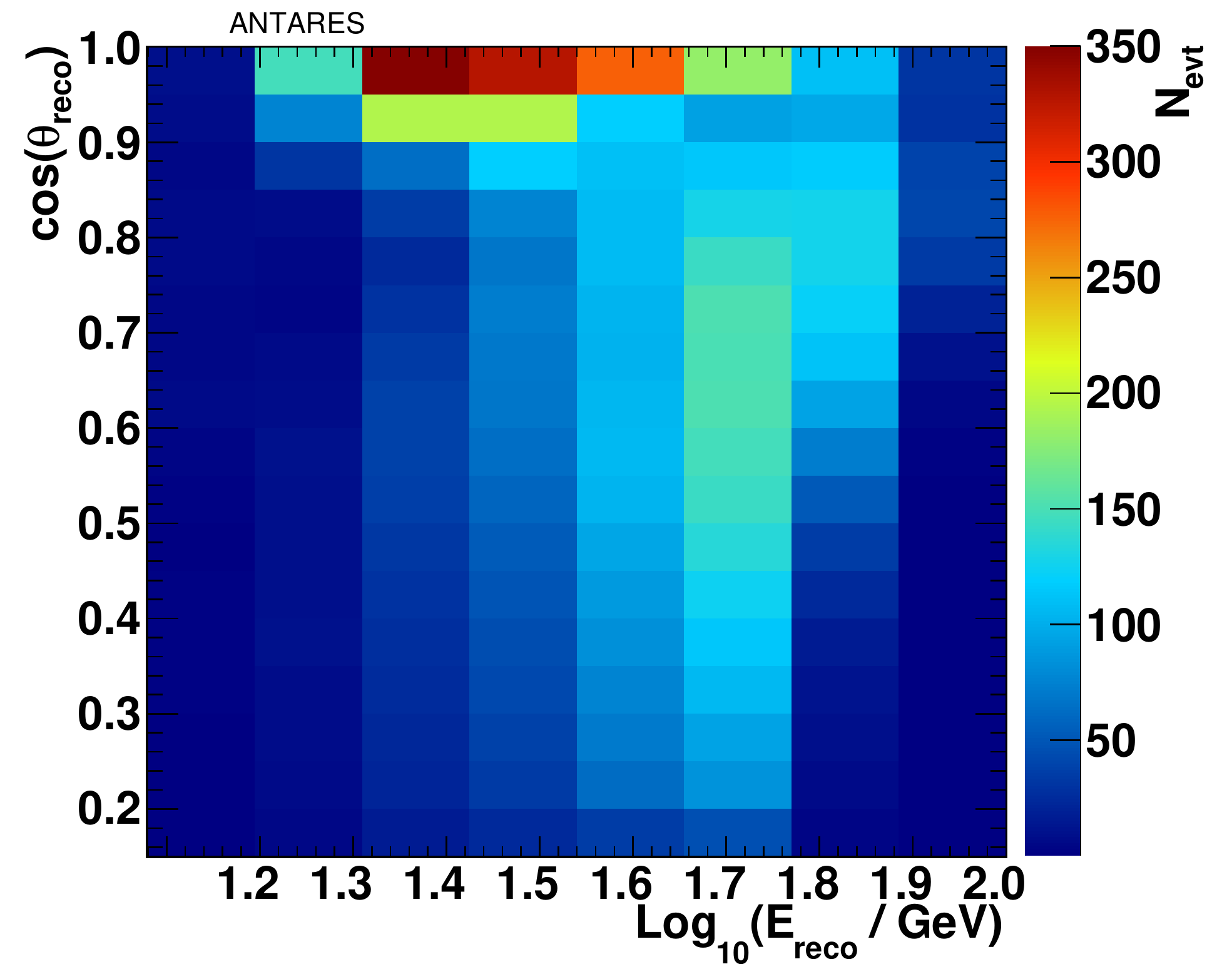}
\end{minipage}
\begin{minipage}[h!]{0.49\linewidth}
\centering
\includegraphics[width=\linewidth]{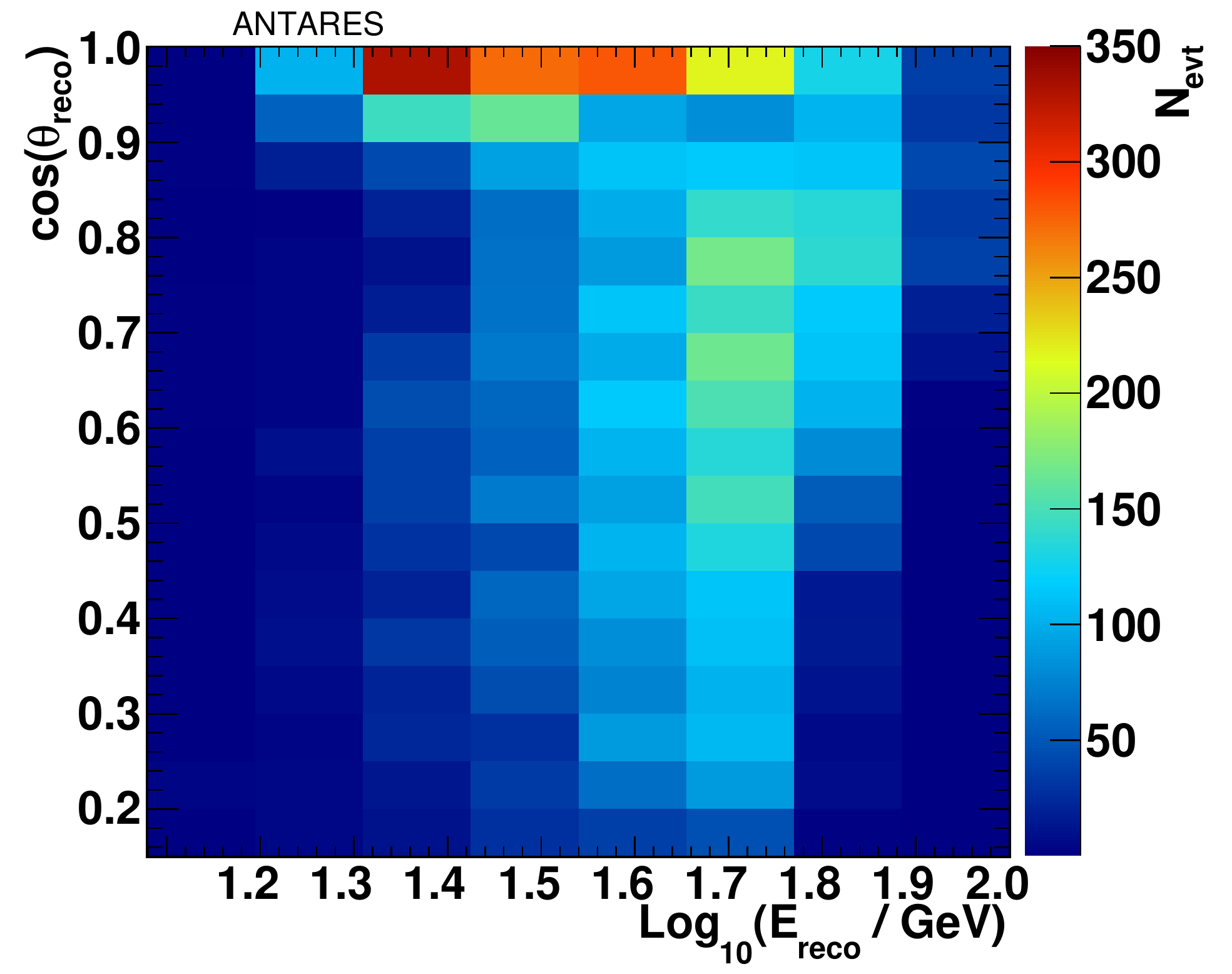}
\end{minipage}
\caption{Number (colour scale on the right side) of selected MC events assuming no oscillation (left panel) and selected data (right panel), binned according to the logarithm of the reconstructed energy, $\log_{10}(E_{\mathrm{reco}}/\mathrm{GeV})$, and the reconstructed cosine of zenith, $\cos\theta_{\mathrm{reco}}$. The first energy bin contains all events with $\log_{10}(E_{\mathrm{reco}}/\mathrm{GeV}) < 1.2$.}
\label{fig:3}
\end{figure}

After applying the event selection criteria described above on the data sample, a total of 7710 events have been selected, 1950 from method  $\mathcal{A}$ (SL), 3682 from method  $\mathcal{A}$ (ML) and 2078 from method  $\mathcal{B}$. In Figure~\ref{fig:3} the event distribution as a function of the logarithm of the reconstructed energy, $\log_{10}(E_{\mathrm{reco}}/\mathrm{GeV})$, and the cosine of the reconstructed zenith angle, $\cos\theta_{\mathrm{reco}}$, is shown. The distribution of the MC expectation assuming no neutrino oscillation (left panel) is compared to what is observed in data (right panel). Eight bins in $\log_{10}(E_{\mathrm{reco}}/\mathrm{GeV})$ have been considered, seven from 1.2 to 2.0, plus an additional underflow bin which accounts for all events with $\log_{10}(E_{\mathrm{reco}}/\mathrm{GeV}) < 1.2$; there are 17 bins in $\cos\theta_{\mathrm{\mathrm{reco}}}$, from 0.15 to 1.0, the latter denoting vertically up-going events.

%\indent In order to check the agreement between the selected data sample and the corresponding MC one, detailed data/MC comparisons have been performed. In Figure~\ref{fig:6b}, as an example, the distribution of the reconstructed cosine of the zenith angle, $\cos\theta_{reco}$ is shown. The MC events have been weighted both assuming no oscillations and a simplistic 2-flavour oscillation scenario with $\theta_{23} = 41.38^{\circ}$ and $\Delta m^2_{32}=2.46\times10^{-3}$\,eV$^2$~\cite{GlFit}. The ratio data/MC in the lower panel is shown using the MC without oscillations.  

%\newpage
%\begin{figure}[h!]
%\centering
%\includegraphics[width=\linewidth]{Fig6b.pdf}
%\vspace*{-1.7cm}
%\caption{Data/MC comparison for the reconstructed cosine of the zenith angle of selected events. MC events have been weighted both assuming no oscillations (red), and assuming a simple 2-flavour oscillation scenario, with $\Delta m^2_{32}~=~2.46~\times~10^{-3}$\,eV$^2$ and $\theta_{23}=41.38^{\circ}$ (green). The Data/MC ratio is computed using MC without oscillations.}
%\label{fig:6b}
%\end{figure} 

The final fit has been performed on the 2-dimensional histograms shown in Figure~\ref{fig:3}. The fit follows a log-likelihood approach, by minimising the function:
\begin{equation}
-2 \log \mathcal{L} = 2\sum_{i,j} [N^{MC}_{i,j}(\bar{p},\bar{\eta}) - N^{data}_{i,j}\cdot \log N^{MC}_{i,j}(\bar{p},\bar{\eta})] +  \sum_k \frac{(\eta_k - <\eta_k>)^2}{\sigma^2_{\eta_k}},
\label{eq:3}
\end{equation}
where the first sum runs over the histogram bins of $\log_{10}(E_{\mathrm{reco}}/\mathrm{GeV})$ and $\cos\theta_{\mathrm{reco}}$, $N^{data}_{i,j}$ is the number of events in bin \textit{(i,j)} and $N^{MC}_{i,j}(\bar{p}, \bar{\eta})$ the corresponding number of expected MC events in the same bin. This number depends on the set of oscillation parameters, $\bar{p}$, that are under investigation, as well as on the set of parameters related to systematic uncertainties, $\bar{\eta}$, as described in the next subsections. 
\textcolor{black}{The dependency on oscillation parameters is taken into acount for CC interactions of all neutrino flavours which contribute to the final event sample}.
The second sum runs over the number of nuisance parameters taken into account, $<\eta_k>$ being the assumed prior of the parameter \textit{k}, and $\sigma_{\eta_k}$ its uncertainty. The log-likelihood function converges to the standard $\chi^2$ for bins with high statistics. For bins with a small number of entries the log-likelihood is more adequate.

Since the treatment of the systematic uncertainties slightly differs between the standard atmospheric oscillation analysis and the sterile neutrino analysis, they are described separately in the following subsections.

\subsection{Treatment of systematics for the standard oscillation analysis}
\label{subsec:5.1}
The standard oscillation analysis accounts for six sources of systematic uncertainties. Three are related to the atmospheric neutrino flux. A global neutrino normalisation factor, $n_{\nu}$, which is left unconstrained during the fit, accounts for uncertainties on the total number of expected events. A variation $\Delta \gamma$ in the nominal neutrino flux spectral index has been used as additional nuisance parameter. Uncertainties on the neutrino/anti-neutrino flux ratio, $\nu/\bar{\nu}$, and on the flux asymmetry between up-going and horizontal neutrinos, $\nu_{\mathrm{up}}/\nu_{\mathrm{hor}}$, have also been taken into account. These uncertainties~\cite{Barr} have been parametrised by the IceCube Collaboration~\cite{ICOsc}. Such parameterisations compute a correction on the number of expected events as a function of the neutrino energy, flavour, chirality, direction and the value of the uncertainty on the flux ratio. The two ratios considered in this analysis have been found to be strongly correlated, thus a unique nuisance parameter is considered in the fit.

An additional source of systematic uncertainty is the limited knowledge of the neutrino interaction model. At the energy of interest for this study, the cross section is dominated by deep inelastic scattering (DIS), with a smaller contribution from quasi elastic (QE) and resonant (RES) scattering. Uncertainties in the DIS cross section can be incorporated in the global flux normalisation factor $n_{\nu}$, as well as in the correction to the spectral index $\Delta\gamma$. For what concerns the QE and RES processes, dedicated studies have been performed with gSeaGen~\cite{gSeaGen}, which uses GENIE~\cite{GENIE} to model neutrino interactions. The dominant systematic is found to be related to the axial mass for CC resonance neutrino production, $M_{A}$. Its default value is 1.12~$\pm$~0.22\,GeV~\cite{GENIE}. By varying this parameter by $\pm 1\sigma$, the correction with respect to the expected number of events has been computed as a function of the true neutrino energy and this parameterisation is used in the final fit.

Apart from the oscillation parameters under investigation, $\Delta m^2_{32}$ and $\theta_{23}$, the other oscillation parameters may play a role, but their effect is limited for this study. In particular, $\theta_{13}$ is left free in the fit 
\textcolor{black}{but treated with a Gaussian prior at $\theta_{13} = (8.41 \pm 0.28)^\circ$, which is taken from a global fit~\cite{GlFit}} as well as the values of the solar neutrino parameters, which are kept fixed: $\Delta m^2_{21} = 7.37\times 10^{-5}$~\,eV$^2$ and $\sin^2\theta_{12}=0.297$. Different values of $\delta_{CP}$ have been tested at the stage of the MC sensitivity study and found to have no impact on the final result. Therefore $\delta_{CP}$ is fixed at zero.

The number of atmospheric muons, $N_\mu$, contaminating the neutrino sample, is treated as an additional nuisance parameter. Its value and uncertainty, determined with the data-driven technique, are used as a prior.

Finally, detector and sea water related systematics have been studied as well.
% with uncertainties from~\cite{AGUILAR2010179}. 
 Dedicated MC simulations have been generated with modified OM photon detection efficiencies and a modified water absorption length, assuming a variation of $\pm 10\%$ from the nominal value, but keeping the same wavelength dependence. 
\textcolor{black}{
The overall OM efficiency can be easily adjusted to the measured coincidence rates 
from $^{40}$K decays~\cite{K40Pap} which makes the chosen 10\% variations a conservative benchmark value, in line with early studies performed on ANTARES OMs~\cite{Hamamatsu}. The water absorption length had been measured several times at the ANTARES site~\cite{AbsL}. The different measurements, taken at two different wavelengths, vary within about 10\%.}
 
\begin{figure}[htbp]
\centering
\includegraphics[width=0.8\linewidth]{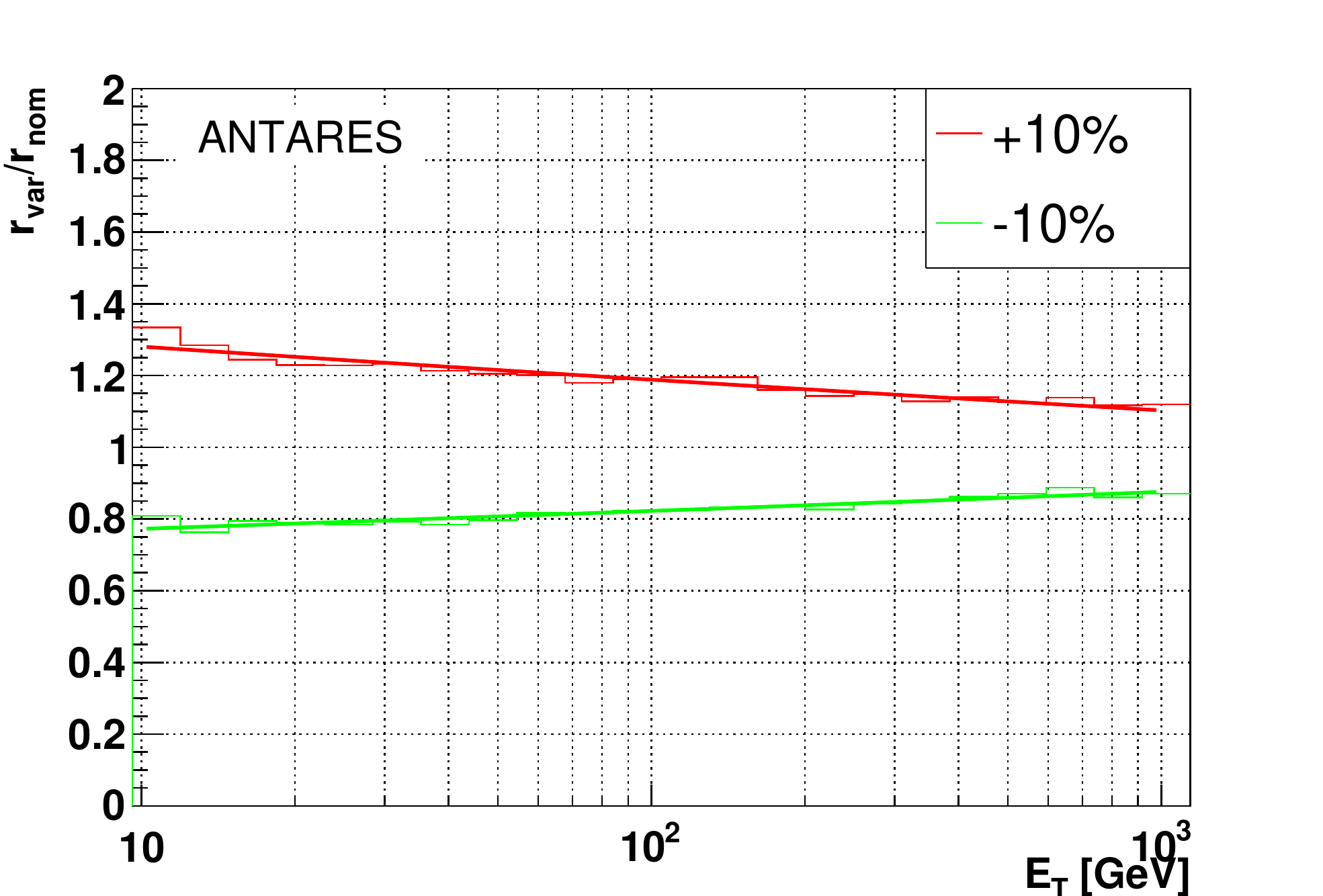}
\caption{Expected event ratios for $\nu_{\mu}$ CC events, as a function of true neutrino energy, due to a $+10$\% (red) and $-10$\% (green) variation from the nominal value of the OM photon detection efficiency.}
\label{fig:4}
\end{figure}

The correction to the event rates, obtained by dividing the event rates from the modified MC simulation ($r_{\mathrm{var}}$)  and the one from the nominal MC simulation ($r_{\mathrm{nom}}$), has been computed as a function of the MC neutrino energy and zenith angle for $\nu_{\mu}$ CC events, reconstructed as up-going. While no zenith-dependent effect is seen, the energy response of the detector is affected by these variations. The resulting distributions have been fitted, in the energy range $10-10^3$\,GeV, with a function of the form:
\begin{equation}
f_\epsilon(\mathrm{E_T}) = A_\epsilon\cdot (\mathrm{E_T}/\mathrm{E_0})^{B_\epsilon},
\label{eq:6.2}
\end{equation}
where $\mathrm{E_T}$ is the MC true neutrino energy, $A_\epsilon$, $B_\epsilon$ are the two fitted parameters describing the effect of the modified OM photon detection efficiencies and $E_0 = 100$~GeV defines the reference energy for $A_\epsilon$. Figure~\ref{fig:4} shows the distribution of the event ratios as a function of true neutrino energy, together with its parameterisation.

The effect of the modified water absorption length is described by the same functional form of Eq.~\ref{eq:6.2} using $A_w$ and $B_w$ as the corresponding fit parameters. The values of the fitted parameters $A_\epsilon, B_\epsilon, A_w$ and $B_w$ are listed in Table~\ref{tab:2}.
The effects of $A_\epsilon$ and $A_w$ are taken into account in the minimisation procedure by the global normalisation factor, $n_{\nu}$, which is left unconstrained, while $B_\epsilon$ and $B_w$ are covered by the uncertainty of the prior on the spectral index, $\Delta\gamma$ (see Table~\ref{tab:4}).
\begin{table}[htbp]
\centering
\begin{tabular}{|c|c|c|c|c|}
\hline
  &  $A_\epsilon$ & $B_\epsilon$ & $A_w$ & $B_w$ \\
\hline
$+10\%$ & 1.19 & -0.03 & 1.16 & -0.02\\
$-10\%$ & 0.82 & 0.03 & 0.92 & 0.02\\
\hline
\end{tabular}
\vspace*{0.1cm}
\caption{Fitted values for the parameterisation of the event weight correction with a variation of $\pm 10\%$ from the nominal value of the OM photon detection efficiency and water absorption length.}
\label{tab:2}
\end{table}

\subsection{Treatment of systematics for the sterile oscillation analysis}
\label{subsec:5.2}
For the sterile analysis, the flux as well as the cross section related systematic uncertainties are treated in the same way as described in the previous subsection.

Since the effect of a sterile neutrino would modify the oscillation pattern in a similar way as $\Delta m^2_{32}$ and $\theta_{23}$ do, these parameters are considered to be one of the sources of systematic uncertainty for this analysis. 
\textcolor{black}{Both $\Delta m^2_{32}$ and $\theta_{23}$ are} left unconstrained as recommended in~\cite{OctSt}. The other standard oscillation parameters are treated as previously discussed.

As discussed in Section~\ref{sec:1}, the addition of a sterile neutrino in the model implies six new mixing parameters to be accounted for. 
The mixing angle $\theta_{14}$ and its associated phase $\delta_{14}$ have been fixed at zero, since they mainly affect the $\nu_e$ channel.
The fast oscillations due to $\Delta m^2_{41} \gtrsim 0.5$\,eV$^2$ are unobservable due to the limited energy resolution of the detector, making $\Delta m^2_{41}$ not measurable. 
%Its precise value has no impact on the analysis as long as it is large compared to $\Delta m^2_{32}$. For this reason, 
It has been kept fixed at 0.5\,eV$^2$. 
\textcolor{black}{The choice of the neutrino mass hierarchy (NMH) as well as $\delta_{24}$ are expected to impact the result. Therefore both normal and inverted hierarchy (NH/IH) and various values of $\delta_{24}$ have been tested during the fit.}
%$\delta_{24}$, instead, has been found to have an effect on the sensitivity to the mixing angles to be constrained, therefore it has been left free during the fit.
Furthermore, to ensure the stability of the fit procedure, the atmospheric muon contamination has been fixed at the value found by the standard oscillation analysis. It has been verified that this choice does not lead to better constraints with respect to the case of a free muon contamination.

\section{Results}
\label{sec:6}
The minimisation procedure has been done using the ROOT package Minuit2~\cite{ROOT}, applied to the function introduced in Equation~\ref{eq:3}. Results are presented in the following subsections, for the standard oscillation analysis and the sterile oscillation analysis, respectively.

\subsection{Results for the standard oscillation analysis}
\label{subsec:6.1}
In Table~\ref{tab:4} the complete list of all the fitted parameters for the standard oscillation analysis is shown, together with their best-fit values and their priors. \textcolor{black}{Due to the high energy threshold of ANTARES this analysis is not sensitive to the NMH. The results hold for both NH and IH}.
The best-fit value is found for $\Delta m^2_{32}$ at $(2.0^{+0.4}_{-0.3})\times 10^{-3}$\,eV$^2$, which is compatible with the current world best-fit value~\cite{GlFit2}. The mixing angle $\theta_{23}$ is found to be compatible with maximal mixing within its error. The global normalisation factor for neutrinos, $n_{\nu}$, is found to be 18\% lower. This value is within the atmospheric neutrino flux uncertainties and it is compatible with what was reported by other analyses~\cite{ICOsc}. A non-negligible pull is found on $\nu/\bar{\nu}$. This parameter seems to compensate for the low value of $n_{\nu}$:
\textcolor{black}{this has been derived from an alternative fit, for which all nuisance parameters but $n_{\nu}$ have been fixed, to allow a more direct comparison with the result reported in~\cite{AntOsc}. Under these conditions $n_{\nu} = 1.04 \pm 0.02$ is found.}
Concerning the spectral index correction, $\Delta\gamma$, no significant distortion from the nominal value is observed. The fitted value for the atmospheric muon contamination shows a strong pull and it is found incidentally close to the MC expectations.
For both $\theta_{13}$ and $M_{A}$ the best fit values and their errors are found at the corresponding prior, which indicates no sensitivity to these parameters.  
{\textcolor{black}{This can be understood as the $\nu_\mu$ survival probability does not depend on $\sin\theta_{13}$ but only on $\cos\theta_{13}=0.99$ (see Eq.~\ref{eq:1})  whereas  $M_{A}$ mainly affects neutrinos with energies below the detection threshold of ANTARES.}

\begin{table}[htbp]
\centering
\begin{tabular}{|c|c|c|}
\hline
Parameter                             & Prior           & Fit result\\
\hline
$\Delta m^2_{32}$ [10$^{-3}$\,eV$^2$] & none            & $2.0^{+0.4}_{-0.3}$\\
$\theta_{23}$ [$^{\circ}$]            & none            & $45^{+12}_{-11}$\\
$n_{\nu}$                             & none            & $0.81^{+0.10}_{-0.09}$\\%0.82    $\pm$ 0.09 \\
$\nu/\overline{\nu}$ [$\sigma$]       & 0.0  $\pm$ 1.0  & $1.10^{+0.64}_{-0.56}$\\%1.1     $\pm$ 0.6\\
$\Delta \gamma$                       & 0.00 $\pm$ 0.05 & --0.003 $\pm$ 0.036\\
$N_{\mu}$                             & 740  $\pm$ 120  & $414^{+48}_{-24}$\\ %+:fudge factor 1.788 = sqrt(24/7.55)  %415 $\pm$ 22\\ 
$\theta_{13}$ [$^{\circ}$]            & 8.41 $\pm$ 0.28 & 8.41    $\pm$ 0.28\\
$M_{A}$ [$\sigma$]                    & 0.0  $\pm$ 1.0  & 0.0     $\pm$ 1.0\\
\hline
\end{tabular}
\caption{Priors and fitted values obtained from the minimisation for all the parameters considered in the standard oscillation analysis.}
\label{tab:4}
\end{table}

The distribution of the ratio between the reconstructed energy and the cosine of the reconstructed zenith is shown in Figure~\ref{fig:6}. This ratio is affected by the oscillation phenomenon as can be seen for the lowest values of $E_{\mathrm{reco}}/\cos\theta_{\mathrm{reco}}$. For comparison, also the distribution of MC assuming no neutrino oscillation, as well as the one assuming the world best-fit values~\cite{GlFit2} are shown. \textcolor{black}{The latter two are calculated with all nuisance parameters at their nominal values}. Such a 1D distribution does not carry the full information exploited in the fit, which is performed on the 2D distribution shown in Figure~\ref{fig:3}. 
\textcolor{black}{While compatible with world data, ANTARES results seem to prefer a somewhat shallower (or energy shifted) oscillation minimum.}

\begin{figure}[htbp]
\begin{minipage}[h!]{0.49\linewidth}
\centering
\includegraphics[width=\linewidth]{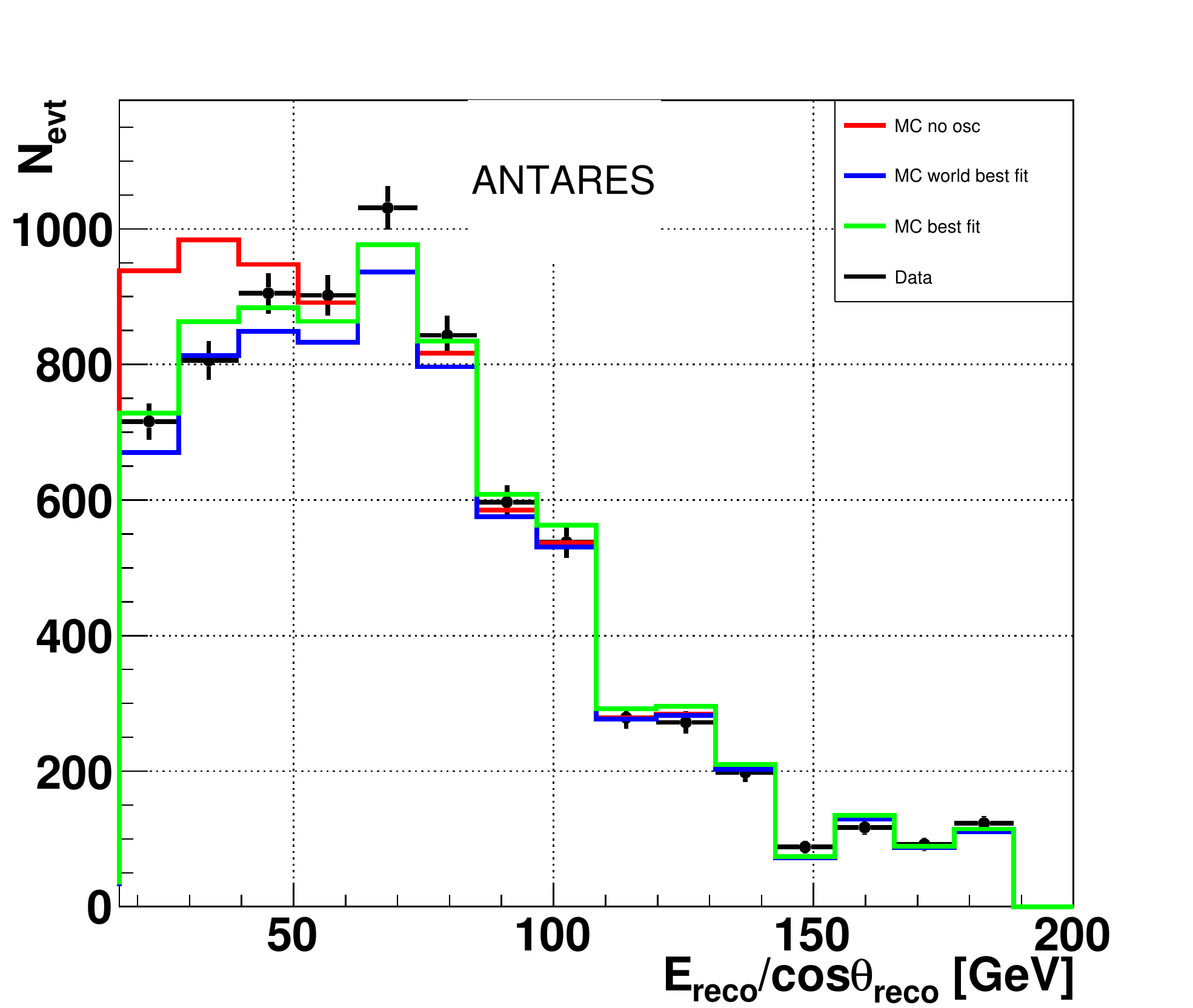}
\end{minipage}
\begin{minipage}[h!]{0.49\linewidth}
\centering
\includegraphics[width=\linewidth]{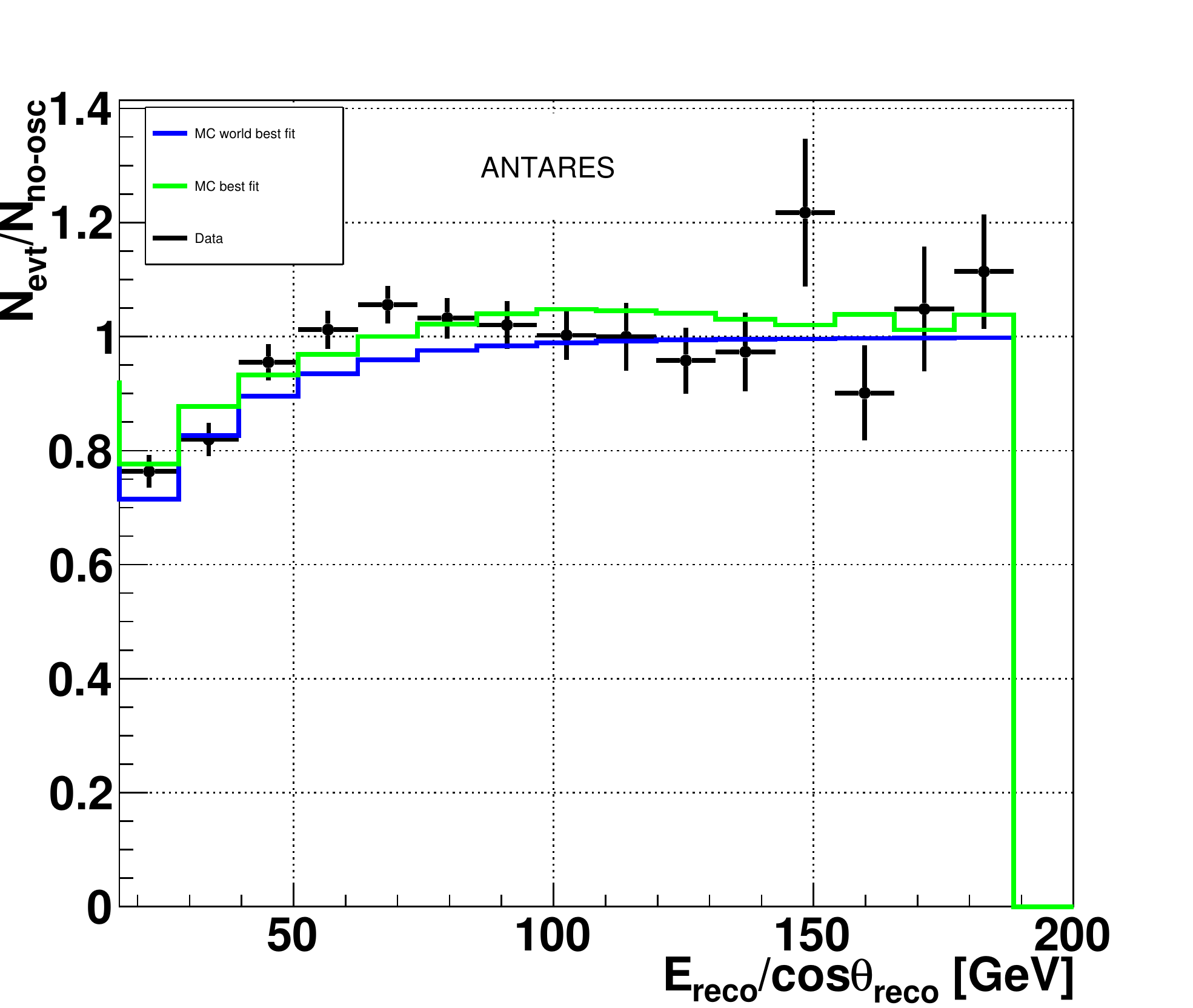}
\end{minipage}
\caption{$E_{\mathrm{reco}}/\cos\theta_{\mathrm{reco}}$ distribution for data (black), MC without oscillation (red), MC assuming the world best-fit values (blue)~\cite{GlFit2} and MC assuming best-fit values of this analysis (green). \textcolor{black}{The left plot shows event numbers while the right plot illustrates the event ratio with respect to the MC without oscillations.}}
\label{fig:6}
\end{figure} 

In Figure~\ref{fig:7} the 90\% CL contour obtained in this work, in the plane of $\sin^2\theta_{23}$ and $\Delta m^2_{32}$, is compared to those published by other experiments. The 1D projections, \textcolor{black}{after profiling over the other variable,} are shown as well. Confidence level contours have been computed by looping over a fine grid of values in $\Delta m^2_{32}$ and $\theta_{23}$ and minimising the negative log-likelihood over all the other parameters.

\begin{figure}[htbp]
\centering
\includegraphics[width=0.8\linewidth]{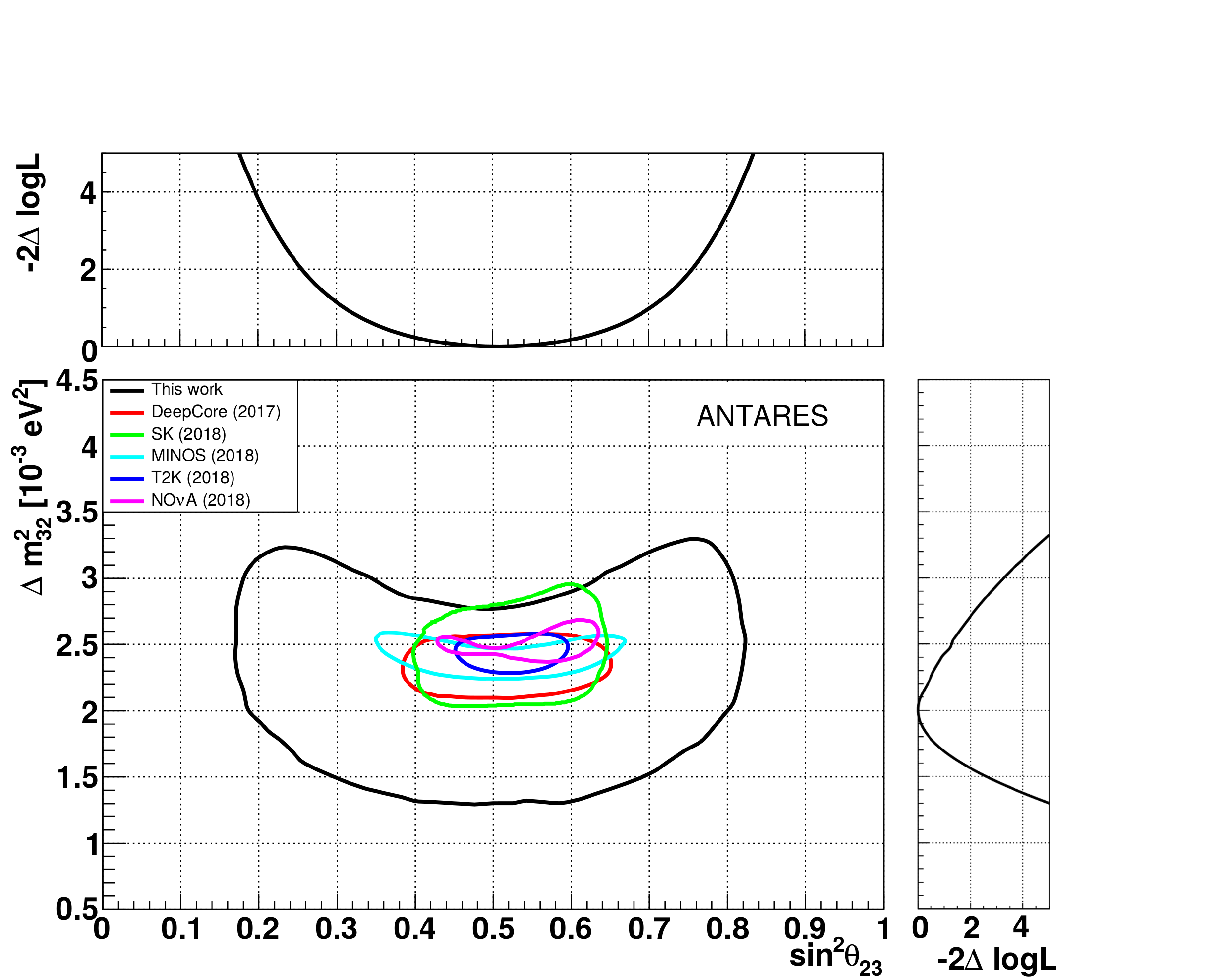}
\caption{Contour at 90\% CL in the plane of $\sin^2\theta_{23}$ and $\Delta m^2_{32}$ obtained in this work (black line) and compared to the results by other experiments: IceCube/DeepCore (red)~\cite{ICOsc}, Super-Kamiokande (green)~\cite{SKOscNu18}, NO$\nu$A (purple)~\cite{NOvANu18}, T2K (blue)~\cite{T2KNu18}, and MINOS (light blue)~\cite{MINOSNu18}. The lateral plots show the 1D projections on the plane of the two oscillation parameters under study.}
\label{fig:7}
\end{figure} 

The non-oscillation hypothesis has been tested by performing the minimisation with a fixed null value of the oscillation parameters, and it is discarded with a significance of $4.6\sigma$, compared to $2.3\sigma$ in our previous analysis~\cite{AntOsc}.

\subsection{Results for the sterile oscillation analysis}
\label{subsec:6.2}
In Table~\ref{tab:5} the complete list of all the fitted parameters for the sterile oscillation analysis \textcolor{black}{for NH and IH} is shown, together with their best-fit values and their priors. While $\theta_{24}$ is found to be compatible with zero, the best fit for $\theta_{34}$ is found at a non-zero value. 
\textcolor{black}{This can be understood from the slight preference of the ANTARES data for a shallower oscillation dip (see discussion related to Fig.~\ref{fig:6}), 
which can be easily provided by a non-zero value of $\sin\theta_{34}$ (see Fig.~\ref{fig:Sterile}).}
%The errors reported in the table are evaluated from the fitting procedure assuming a parabolic shape of the negative likelihood distribution around the minimum.
\textcolor{black}{The non-sterile hypothesis is found at $-2\Delta\log\mathcal{L}=4.4$ which corresponds to a 2-parameter p-value of 11\%.}
%It can be seen from Figure~\ref{fig:9} (\textcolor{black}{projection on $\sin^2\theta_{34}\cos^2\theta_{24}$ after profiling over $\sin^2\theta_{24}$}) that the fitted value of $\theta_{34}$ is compatible with 0 at the level of 2.2$\sigma$. 
The fitted values of $\Delta m^2_{32}$ and $\theta_{23}$ are slightly different but consistent with respect to the ones obtained in the standard oscillation analysis. 
%This is due to the fact that here a prior on $\Delta m^2_{32}$ is applied, which in turn affects the fitted value for the mixing angle. 
\textcolor{black}{The complex phase $\delta_{24}$ is found at $180^\circ$. For IH instead the fit prefers $\delta_{24}=0^\circ$ with otherwise identical results, as expected from the degeneracy between NMH and $\delta_{24}$ (see Appendix)}.
%A large uncertainty is found for $\delta_{24}$, \textcolor{black}{as its influence disappears with vanishing $\theta_{24}$ (see Appendix).}
For the other parameters a similar behaviour as for the standard oscillation analysis is observed.

\begin{table}[htbp]
\centering
\begin{tabular}{|c|c|c|c|}
\hline
Parameter                             & Prior           & Fit NH                 & Fit IH \\
\hline
$\theta_{24}$ [$^{\circ}$]            & none            & $1.5^{+2.0}_{-5.0}$    & $1.5^{+2.0}_{-5.0}$   \\%0.9     $\pm$ 1.8 \\
$\theta_{34}$ [$^{\circ}$]            & none            & $25.9 ^{+5.1}_{-4.2}$  & $25.9 ^{+5.1}_{-4.2}$  \\%24      $\pm$ 4\\
$\delta_{24}$ [$^{\circ}$]            & none            & $180\pm 71$            & $0 \pm 72$             \\%0       $\pm$ 120 \\
$n_{\nu}$                             & none            & $0.84^{+0.10}_{-0.09}$ & $0.84^{+0.10}_{-0.09}$ \\%0.81    $\pm$ 0.09 \\
$\nu/\overline{\nu}$ [$\sigma$]       & 0.0  $\pm$ 1.0  & $1.07^{+0.63}_{-0.55}$ & $1.07^{+0.63}_{-0.55}$ \\%1.1     $\pm$ 0.6 \\
$\Delta\gamma$                        & 0.00 $\pm$ 0.05 & $-0.011 \pm 0.036$     & $-0.011 \pm 0.036$     \\ %--0.001 $\pm$ 0.035 \\
$\Delta m^2_{32}$ [$10^{-3}$\,eV$^2$] & none            & $3.0^{+0.8}_{-0.6}$    & $-3.0^{+0.6}_{-0.8}$   \\ %2.49    $\pm$ 0.13\\
$\theta_{23}$ [$^{\circ}$]            & none            & $52 \pm 8$             & $52 \pm 8$             \\%49      $\pm$ 7 \\
$\theta_{13}$ [$^{\circ}$]            & 8.41 $\pm$ 0.28 & $8.41 \pm 0.28$        & $8.41 \pm 0.28$        \\%8.41    $\pm$ 0.28 \\
$M_{A}$ [$\sigma$]                    & 0.0  $\pm$ 1.0  & $0.11^{+0.93}_{-0.97}$ & $0.11^{+0.93}_{-0.97}$ \\%0.1     $\pm$ 1.0 \\
\hline
\end{tabular}
\caption{Priors and fitted values obtained from the minimisation for all the parameters considered in the sterile oscillation analysis.} 
%$\theta_{34}$ is compatible with zero at a 1-parameter p-value of 3.6\%.}
\label{tab:5}
\end{table} 
%\begin{figure}[htbp]
%\centering
%\includegraphics[width=\linewidth]{Fig8.pdf}
%\caption{99\% CL exclusion regions and best fit points for the 3+1 neutrino model in the parameter space of $|U_{\mu 4}|^2=\sin^2\theta_{24}$ and $|U_{\tau 4}|^2=\sin^2\theta_{34}\cos^2\theta_{24}$ obtained in this work (black) and compared to the ones published by IceCube/DeepCore~\cite{ICStLow} (red) and Super-Kamiokande~\cite{SKSt} (green). The 1D projections are also shown for the results of the work.}
%\label{fig:8}
%\end{figure} 

%The non-sterile hypothesis is found to be slightly disfavoured, similar to what is observed in the other analyses~\cite{ICStLow,SKSt}. 
\textcolor{black}{Exclusion contours are built by applying Wilks' theorem.}
In Figure~\ref{fig:9} the resulting  90\% and 99\% CL exclusion limits have been computed on a 2D grid in the plane of the two matrix elements, namely $|U_{\mu 4}|^2 = \sin^2\theta_{24}$ and $|U_{\tau 4}|^2 = \sin^2\theta_{34}\cos^2\theta_{24}$. \textcolor{black}{The exclusion limit for unconstrained $\delta_{24}$, which corresponds to both [NH,$\delta_{24}=180^\circ$] or [IH,$\delta_{24}=0^\circ$], can be directly compared to the IceCube/DeepCore~\cite{ICStLow} (IH) limit. Also shown are limits for NH and $\delta_{CP}=0^\circ$ which allow a direct comparison with the results from IceCube/DeepCore~\cite{ICStLow} (NH) and Super-Kamiokande~\cite{SKSt}.} All three experiments find the best fit for $|U_{\tau 4}|^2$ to differ from zero. 
\textcolor{black}{Our results exclude regions of the parameter space not yet excluded by other experiments.}
%In some regions of the plane, ANTARES limits are more stringent. 

%It is worth mentioning that both the energy range and the systematic treatment, in particular concerning the standard atmospheric oscillation parameters, are different among the three results illustrated in the figure. 
The IceCube/DeepCore analysis~\cite{ICStLow} is limited to events with reconstructed energy lower than 56~GeV, while the distortion on the oscillation pattern possibly produced by the presence of a sterile neutrino would be evident also at higher reconstructed energies. The present analysis includes events with reconstructed energy up to 100 GeV.
\textcolor{black}{It has been verified that the ANTARES limits degrade when restricting the analysis to events with $E_\mathrm{reco}<56$~GeV.}
\textcolor{black}{In this work both of the standard atmospheric oscillation parameters $\Delta m^2_{32}$ and $\sin^2(2\theta_{23})$ are left unconstrained in line with the IceCube/DeepCore analysis~\cite{ICStLow}.} 
%The Super-Kamiokande analysis~\cite{SKSt} applies priors both on $\Delta m^2_{32}$ and on $\sin^2(2\theta_{23})$. %which could partially explain the more stringent limit obtained at low values of $\theta_{24}$. 
%Finally, this analysis let $\delta_{24}$ completely unconstrained, while the other analyses fix this parameter at zero. 

\begin{figure}[htbp]
\centering
\includegraphics[width=\linewidth]{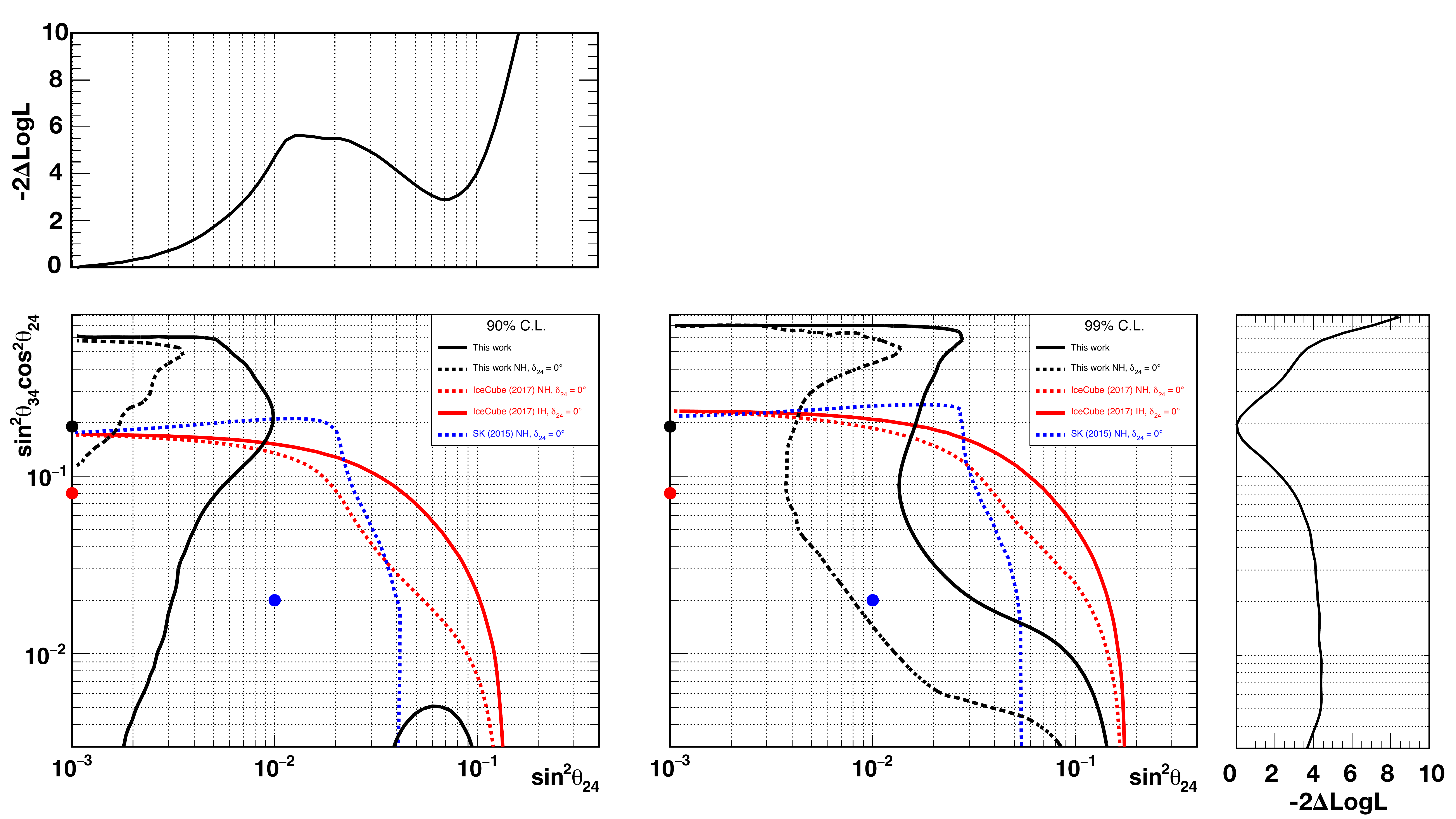}
\caption{90\% (left) and 99\% (right) CL limits for the 3+1 neutrino model in the parameter plane of $|U_{\mu 4}|^2=\sin^2\theta_{24}$ and $|U_{\tau 4}|^2=\sin^2\theta_{34}\cos^2\theta_{24}$ obtained in this work (black lines), and compared to the ones published by IceCube/DeepCore~\cite{ICStLow} (red) and Super-Kamiokande~\cite{SKSt} (blue). \textcolor{black}{The dashed lines are obtained for NH and $\delta_{24}=0^\circ$ while the solid lines are for an unconstrained $\delta_{24}$ (this work) or for IH and $\delta_{24}=0^\circ$ (IceCube/Deepcore) respectively.} The coulored markers indicate the best-fit values for each experiment. 
The 1D projections \textcolor{black}{after profiling over the other variable} are also shown for the result of this work.}
\label{fig:9}
\end{figure} 

%From the 1D projections of Figure~\ref{fig:9}, 
\textcolor{black}{After profiling over the other variable,} the following limits on the two matrix elements can be derived:
\begin{eqnarray}
|U_{\mu 4}|^2  &<& 0.007~(0.13)~ \mathrm{at}~ 90\%~(99\%)~\mathrm{CL},\\ 
|U_{\tau 4}|^2 &<& 0.40~(0.68)~ \mathrm{at}~ 90\%~(99\%)~\mathrm{CL}. 
\end{eqnarray}

\section{Conclusions}
\label{sec:7}
Ten years of ANTARES data have been analysed to provide a measurement of the atmospheric neutrino oscillation parameters. The analysis chain has been optimised with respect to our previously published study, by combining two track reconstruction algorithms and introducing a more elaborate treatment of various sources of systematic uncertainties. The results, $\Delta m^2_{32}= (2.0^{+0.4}_{-0.3})\times 10^{-3}$\,eV$^2$ and $\theta_{23} = (45^{+12}_{-11})^\circ$, are consistent with what has been published by other experiments. The non-oscillation hypothesis is discarded with a significance of $4.6\sigma$.

Exploiting the same analysis chain and the same data set, a further study has allowed to constrain, for the first time with ANTARES, the parameter space of the 3+1 neutrino model, which foresees the existence of one sterile neutrino. 
\textcolor{black}{ANTARES excludes values of the parameter space not yet excluded by other experiments.}
%ANTARES limits are in some regions of the parameter space more stringent than the results reported by other experiments. 

\section{Acknowledgements}
The authors acknowledge the financial support of the funding agencies:
% France:
Centre National de la Recherche Scientifique (CNRS), Commissariat \`a
l'\'ener\-gie atomique et aux \'energies alternatives (CEA),
Commission Europ\'eenne (FEDER fund and Marie Curie Program),
Institut Universitaire de France (IUF), IdEx program and UnivEarthS
Labex program at Sorbonne Paris Cit\'e (ANR-10-LABX-0023 and
ANR-11-IDEX-0005-02), Labex OCEVU (ANR-11-LABX-0060) and the
A*MIDEX project (ANR-11-IDEX-0001-02),
R\'egion \^Ile-de-France (DIM-ACAV), R\'egion
Alsace (contrat CPER), R\'egion Provence-Alpes-C\^ote d'Azur,
D\'e\-par\-tement du Var and Ville de La
Seyne-sur-Mer, France;
% Germany: 
Bundesministerium f\"ur Bildung und Forschung
(BMBF), Germany; 
% Italy
Istituto Nazionale di Fisica Nucleare (INFN), Italy;
% Netherlands
Nederlandse organisatie voor Wetenschappelijk Onderzoek (NWO), the Netherlands;
% Russia
Council of the President of the Russian Federation for young
scientists and leading scientific schools supporting grants, Russia;
% Romania
Executive Unit for Financing Higher Education, Research, Development and Innovation (UEFISCDI), Romania;
% Spain
Mi\-nis\-te\-rio de Econom\'{\i}a y Competitividad (MINECO):
Plan Estatal de Investigaci\'{o}n (refs. FPA2015-65150-C3-1-P, -2-P and -3-P, (MINECO/FEDER)), Severo Ochoa Centre of Excellence and Red Consolider MultiDark (MINECO), and Prometeo and Grisol\'{i}a programs (Generalitat
Valenciana), Spain; 
% Marocco
Ministry of Higher Education, Scientific Research and Professional Training, Morocco.
% A.O.B.:
We also acknowledge the technical support of Ifremer, AIM and Foselev Marine
for the sea operation and the CC-IN2P3 for the computing facilities.
\textcolor{black}{
We would like to thank J. Coelho for enlightening discussions on matter effects for sterile neutrinos.}

\section{Appendix : Sterile neutrinos and matter effects}

\textcolor{black}{
For the analysis presented in this paper, oscillation probabilities are evaluated with the software package OscProb~\cite{OscProb}. However, in this Appendix some common approximations are applied, to derive analytical formulae. These are NOT used for the analysis itself but allow to get a better understanding of the interplay between different parameters. The $\nu_\mu$ survival probability in vacuum in the $3+1$ model can be simplified with the following two hypotheses~\cite{MalSchw,SKSt}: first, it is assumed, that the first generation decouples completely, i.e. $\Delta m^2_{21}=0$ and $\theta_{12} = \theta_{13} = 0$; second, fast wiggles due to oscillations involving $m_4$ are assumed to be unobservable, i.e. $\sin^2(\Delta m^2_{4i} L/4E)=1/2$ for all $i$. This yields
\begin{equation}
P_{\nu_{\mu}\rightarrow\nu_{\mu}} = (1-|U_{\mu 4}|^2)^2 P_{\mu \mu}^{(3)} + |U_{\mu 4}|^4,
\label{appvac}
\end{equation}
with $P_{\mu \mu}^{(3)}$ the $\nu_\mu$ survival probability in the 3-flavour scheme, i.e. without additional sterile neutrinos. Only $|U_{\mu 4}|^2 = \sin^2\theta_{24}$ can be probed in this scheme, which is applied in most accelerator based $\nu_\mu$ disappearance analyses. However, when analysing atmospheric neutrinos, matter effects cannot be neglected. An analytical formalism is developped in Eqs.~4.13-4.25 of~\cite{SKSt}.  In Eq.~4.13, a complex phase is present in the non-diagonal term of the matrix, which is neglected, i.e. set to zero, in subsequent steps. If instead this phase is kept, $\sin2\theta_s$ in Eq.~4.16 acquires an extra term $\exp(-i\delta)$. 
\begin{eqnarray}
\sin2\theta_s  &=& \frac{2\sqrt{|U_{\mu 4}|^2|U_{\tau 4}|^2(1-|U_{\mu 4}|^2-|U_{\tau 4}|^2)}}{(1-|U_{\mu 4}|^2)(|U_{\mu 4}|^2+|U_{\tau 4}|^2)}e^{-i\delta},\\ 
\cos2\theta_s  &=&   \frac{|U_{\tau 4}|^2-|U_{\mu 4}|^2(1-|U_{\mu 4}|^2-|U_{\tau 4}|^2)}{(1-|U_{\mu 4}|^2)(|U_{\mu 4}|^2+|U_{\tau 4}|^2)}. 
\end{eqnarray}
This in turn modifies Eqs.~4.18 and Eq.~4.19:
%\begin{eqnarray}
\begin{equation}
E_m^2=A_{32}^2 + A_s^2 + 2A_{32}A_s(\sin2\theta_{23}|\sin2\theta_s|\cos\delta+\cos2\theta_{23}\cos2\theta_s),\\ 
\end{equation}
\begin{equation}
\sin2\theta_m=\frac{1}{E_m}\sqrt{A_{32}^2\sin^22\theta_{23}+A_s^2|\sin2\theta_s|^2+2A_{32}A_s\sin2\theta_{23}|\sin2\theta_s|\cos\delta}. 
\end{equation}
%\end{eqnarray}
For $\delta=0$ the original expressions from~\cite{SKSt} are reproduced. With $A_{32}=\Delta m^2_{32}/E_\nu$ and $A_s=\frac{\sqrt{2}}{2}G_FN_n(|U_{\mu4}|^2+|U_{\tau 4}|^2)/2$ 
($G_F$ the Fermi constant and $N_n$ the neutron density) the $\nu_\mu$ survival probability in matter is fully defined and can be written equivalently to Eq.~\ref{appvac} (see also Eq.~4.23 of~\cite{SKSt}):
\begin{equation}
P_{\nu_{\mu}\rightarrow\nu_{\mu}} = (1-|U_{\mu 4}|^2)^2 (1-\sin^22\theta_m\sin^2(E_mL)) + |U_{\mu 4}|^4,
\label{appmat}
\end{equation}
which describes well all features shown in Fig.~\ref{fig:Sterile}. The impact of the CP-phase $\delta$ disappears when either $|U_{\mu 4}|^2=0$ or  $|U_{\tau 4}|^2=0$, which leads to $\sin2\theta_s=0$. Further, $\delta\rightarrow\delta+\pi$ is completely degenerate with changing the mass hierarchy, i.e. swapping the sign of $A_{32}$ if either $\cos2\theta_{23}=0$ or $\cos2\theta_{s}=0$. Deviation from maximal mixing in $\theta_{23}$ or from the symmetry between $|U_{\mu 4}|^2$ and  $|U_{\tau 4}|^2$ defining $\theta_s$ breaks this degeneracy. The impact of the neutrino mass hierarchy on the $\nu_\mu$ survival probability in matter had been pointed out already  in~\cite{Razza}, while the influence of complex phases is also discussed in~\cite{BlennowICNOMAD}.}


\bibliographystyle{JHEP}
\bibliography{./Bibliography}

\end{document}